\documentclass[12pt]{article}
\pdfoutput=1
\usepackage{color,amsmath,amssymb,exscale,psfrag,epsfig}
\usepackage{cite,color,url}
\usepackage{pdfpages}
\usepackage{graphicx}
\usepackage{slashed}
\usepackage[utf8]{inputenc}
\usepackage[T1]{fontenc}
\usepackage{xcolor}
\usepackage{accents}
\usepackage{centernot}
\usepackage{enumerate}
\usepackage{subfig}
\usepackage{float}
\usepackage{placeins}
\usepackage{floatpag}

\usepackage[colorlinks=true
,urlcolor=blue
,anchorcolor=blue
,citecolor=blue
,filecolor=blue
,linkcolor=blue
,menucolor=blue
,linktocpage=true
,pdfproducer=medialab
]{hyperref}
\input epsf
\usepackage{lineno}

\def\bea{\begin{eqnarray}}
\def\eea{\end{eqnarray}}

\definecolor{nicered}{rgb}{0.7,0.1,0.1}
\definecolor{nicegreen}{rgb}{0.1,0.5,0.1}
\hypersetup{colorlinks,citecolor= nicegreen,linkcolor= nicered}

\def\be{\begin{equation}}
\def\te{\end{equation}}
\def\ee{\end{equation}}
\def\ba{\begin{eqnarray}}
\def\bea{\begin{eqnarray}}
\def\nn{\nonumber}
\def\tea{\end{eqnarray}}
\def\ea{\end{eqnarray}}
\def\eea{\end{eqnarray}}

\def\bfra{\begin{frame}}
\def\efra{\end{frame}}
\def\al#1\fal{\begin{align}#1\end{align}}

\def\bfra#1\efra{\begin{frame}#1\end{frame}}

\catcode`\@=11
\def\lsim{\mathrel{\mathpalette\@versim<}}
\def\gsim{\mathrel{\mathpalette\@versim>}}
\def\@versim#1#2{\vcenter{\offinterlineskip
\ialign{$\m@th#1\hfil##\hfil$\crcr#2\crcr\sim\crcr } }}
\catcode`\@=12
\catcode`@=12
\begin{document}
\thispagestyle{empty}
\begin{flushright}
ICAS 052/20
\end{flushright}
\vspace{0.1in}
\begin{center}
	{\Large \bf Containing COVID-19 outbreaks using a firewall} \\
\vspace{0.2in}
{\bf Ezequiel Alvarez$^{(a)\dagger}$,
Leandro Da Rold$^{(b)\star}$,
\\

Federico Lamagna$^{(b)\ddag}$,
Manuel Szewc$^{(a)\diamond}$
}
\vspace{0.2in} \\
{\sl $^{(a)}$ International Center for Advanced Studies (ICAS), UNSAM and CONICET,\\
	Campus Miguelete, 25 de Mayo y Francia, (1650) Buenos Aires, Argentina }
\\[1ex]
{\sl $^{(b)}$ Centro At\'omico Bariloche, Instituto Balseiro and CONICET\\
Av.\ Bustillo 9500, 8400, S.\ C.\ de Bariloche, Argentina}
\end{center}
\vspace{0.1in}

\begin{abstract}
	COVID-19 outbreaks have proven to be very difficult to isolate and extinguish before they spread out.  An important reason behind this might be that epidemiological barriers consisting in stopping symptomatic people are likely to fail because of the contagion time before onset, mild cases and/or asymptomatics carriers.   Motivated by these special COVID-19 features, we study a scheme for containing an outbreak in a city that consists in adding an extra firewall block between the outbreak and the rest of the city.  We implement a coupled compartment model with stochastic noise to simulate a localized outbreak that is partially isolated and analyze its evolution with and without firewall for different plausible model parameters. We explore how further improvements could be achieved if the epidemic evolution would trigger policy changes for the flux and/or lock-down in the different blocks. Our results show that a substantial improvement is obtained by merely adding an extra block between the outbreak and the bulk of the city.
\end{abstract}

\vspace*{2mm}
\noindent {\footnotesize E-mail:
{\tt 
$\dagger$ \href{mailto:sequi@unsam.edu.ar}{sequi@unsam.edu.ar},
$\star$ \href{mailto:daroldl@cab.cnea.gov.ar}{daroldl@cab.cnea.gov.ar},
$\ddag$ \href{mailto:federico.lamagna@cab.cnea.gov.ar}{federico.lamagna@cab.cnea.gov.ar},
$\diamond$ \href{mailto:mszewc@unsam.edu.ar}{mszewc@unsam.edu.ar}
}}


\section{Introduction}
\label{section:introduction}
The outbreak of COVID-19 by the end of 2019 in Asia has spread all over the world in a few months, generating a worldwide crisis with yet unknown consequences. 
Currently the number of detected cases in the world is of order 24 million, with more than 800 thousand reported deaths.
Many Asian and European countries have passed the first wave and are currently receiving second waves of the epidemic, whereas most of the other countries are passing the first wave. 
Given the lack of a vaccine most of the national governments have adopted non-pharmaceutical measures to decrease the speed of contagion and protect vulnerable populations, as restriction of mobility of the populations, prohibition of crowded events and isolation of populations with localized outbreaks, while the capacity of health systems are increased.

A huge effort of the scientific community is devoted to characterize the disease. There has also been a lot of effort on modeling the dynamics of the epidemic. See for example Refs.~\cite{ Giordano_2020,CONTRERAS2020109925, alex2020eliminating, TAGLIAZUCCHI2020109923, Mahajan2020.06.21.20136580, oxfordModel} or the helpful reviews of Refs.~\cite{Shinde2020,review} and references therein. Nevertheless, many authors have pointed that forecasts must be considered with great care.

There have also been many proposals to contain and mitigate the epidemic~\cite{Giordano_2020, oxfordModel, oxfordNPI, covidUS, TAGLIAZUCCHI2020109923}. One of the important strategies for mitigating the epidemic has been the implementation of different levels of barriers around a region with an outbreak of COVID-19. The barriers are expected to contain the expansion of the epidemic, while avoiding a hard quarantine in the whole city.

This article is devoted to model barriers and to propose some strategies that could be useful to mitigate the propagation of the epidemic from a localized outbreak on a city to the rest of it. Since in general a population can not be fully isolated, either because elementary services must be delivered to it, or because for any reason individuals could leave it, we include the possibility of permeable barriers, modeling the flux between the different partially isolated sectors of the city. Then we explore, for given fluxes of individuals crossing the barriers, the probability that the epidemic is transmitted from one sector to the other. We consider two particular cases: first a barrier separating a region with an outbreak from the rest of the city, and second the presence of two barriers surrounding the outbreak, with an intermediate region that plays the role of a firewall. We discuss the differences between both situations. We show that the firewall can be used either to reduce the probability of an outbreak in the larger region by fixing the flux, or to allow a slightly larger population flux without increasing the probability of an outbreak in the larger region. We do not attempt to give a quantitative description of the dynamics of the populations, instead we show some qualitative and generic effects, that perhaps could help to mitigate the transmission of an outbreak. Finally, it is worth noticing that we do not consider social and political effects of the barriers, which should be addressed separately if these strategies would be implemented.

The article is organized as follows: in Sec.~\ref{section:model} we define our model and the values of the relevant parameters; in Sec.~\ref{section:two} we implement this model with and without the firewall between an outbreak and a large region and in Sec.~\ref{section:conclusion} we conclude. We leave the more technical details to the Appendix, dealing with both the stochastic model implementation in Appendix~\ref{appendix-fluctuations} and the equations of the coupled model in Appendix~\ref{section:appendix_interconnected}.

\section{The Model}
\label{section:model}

In this section we define the model we use to study the effects of implementing a firewall strategy. We assume that every city block behaves according to the same underlying model which are then coupled through population fluxes. We work with a SIR-like model~\cite{Kuperman_2013}, that can accommodate the particular COVID-19 characteristics within some degree of approximation, see for example Refs.~\cite{ Giordano_2020,CONTRERAS2020109925, TAGLIAZUCCHI2020109923, Mahajan2020.06.21.20136580, oxfordModel} for similar descriptions.  In particular, we define epidemiological barriers as those that allow to pass only people showing no symptoms, and we discriminate between the population that develop symptoms and the one that do not, since the latter can pass between the blocks. 

\subsection{Compartment model}
\begin{figure}[h!]
\centering
\includegraphics[width=0.7\textwidth]{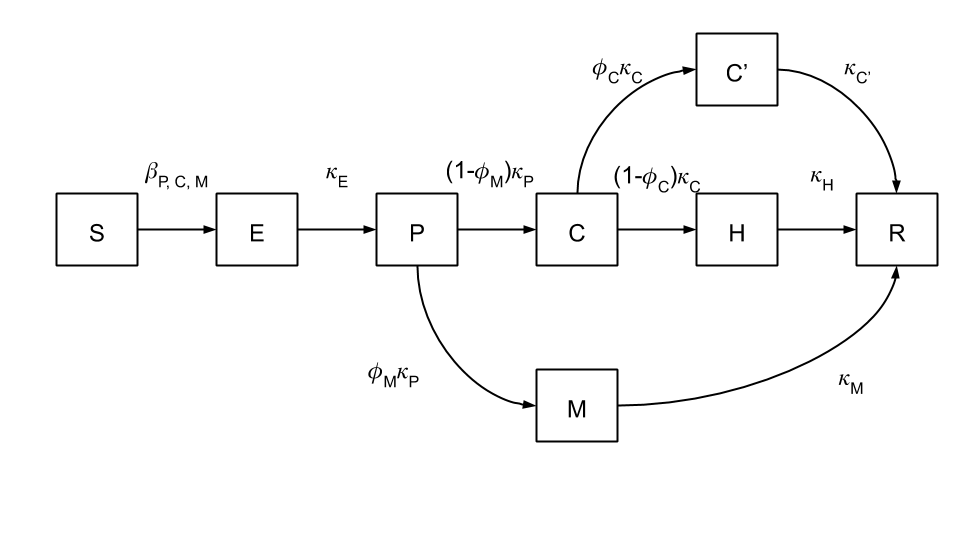}
        \caption{Flow diagram.}
\label{R0}
\end{figure}

On each block, we split our fixed population of $N$ individuals into seven compartments which we call $S, E, P, M, C, C^{'},H,R$. These compartments stand for {\it S}usceptible, {\it E}xposed, {\it P}re-symptomatic, {\it M}ildly symptomatic/asymptomatic, {\it C}ontagious, {\it C'}ontagious and avoiding hospitalization/quarantine, {\it H}ospitalized/quarantined and {\it R}ecovered/dead, respectively.  For the purposes and methods of this work, we consider the group $M$ as those whose level of symptoms is below the threshold to be stopped in an epidemiological barrier.  The population splitting and its interactions are depicted in a flowchart in Fig.~\ref{R0}. 

Interactions between populations are parameterized by $\beta_{P,M,C,C^{'}}$, that encode the disease interactions between a susceptible individual and an infectious one. They depend both on biological and societal factors. For simplicity, we set $\beta_{C}=\beta_{C^{'}}$, meaning that the diagnosed individuals do not change their behavior if they do not to get hospitalized/quarantined.

Once an individual is infected, it advances through one of the possible paths of Fig.~\ref{R0}. The $\kappa_{E,P,M,C,C',H}$ parameters can be thought of as decay rates, meaning that they are the inverse of the characteristic transition times. These characteristic decay rates control the disease duration but do not impose any constraints on the path election. The path choices are controlled through the $\phi_{M,C}$ parameters, which are the fraction of individuals that go from $P$ to $M$ and from $C$ to $C^{'}$. 
In a certain way, these parameters help to encode the different dangers of the COVID-19 epidemic: undiagnosed patients who unknowingly continue to spread the disease and diagnosed individuals who, for whatever reason, continue to spread the disease.

Taking $S,E,P,\dots$ as the number of people in each compartment, the equations describing the evolution of the populations are:~\footnote{By choosing suitable values for the parameters, it is possible to recover a SIR model.}
\begin{equation}
	\begin{aligned}
\frac{d S}{dt} &= - \frac{S}{N-H} \left(\beta_P P + \beta_C (C+C') + \beta_M M \right)\;, & \\
		\frac{d E}{dt} &=  \frac{S}{N-H} \left(\beta_P P + \beta_C (C+C') + \beta_M M \right) -\kappa_{E} E \;, &  \\
\frac{d P}{dt} &= \kappa_E E  - \kappa_P P  \;,  &
\frac{d M}{dt} = \kappa_{P} \phi_M P - \kappa_{M} M \;, \\
\frac{d C}{dt} &= \kappa_{P}(1-\phi_M) P - \kappa_{C} C \;, &
\frac{d H}{dt} = (1-\phi_C) \kappa_{C} C - \kappa_{H} H \;,  \\
\frac{d C'}{dt} &= \phi_C \kappa_{C} C - \kappa_{C'} C'  \;, &
\frac{d R}{dt} = \kappa_{M} M + \kappa_{C'} C' + \kappa_{H} H \;.
	\end{aligned}
\label{slaq_2}
\end{equation}

We do not include injection and extraction of individuals from outside the city, they can be included straightforward but, as described in next section, for the situation we are interested in they are not needed.

During the evolution of the epidemic the $\beta$ parameters are expected to change, responding to the different policies, as well as to societal dynamics. Therefore, we do not expect the model with a set of constant parameters to be able to describe the whole epidemic in a given city. However, since we do not aim to make a forecast, this simple model can accommodate a reasonable qualitative description of the epidemic during a given fraction of time, that captures the core of our proposal.

To solve Eqs.~(\ref{slaq_2}) we implement a numerical method which introduces a daily time-step. Moreover, we treat this model not as deterministic but as a stochastic model by introducing different sources of noise. We detail our implementation in Appendix~\ref{appendix-fluctuations}.

As can be done for SIR models, we define a reproduction number $R_{0}$ that captures the expected new infections per infected individual at initial time:
\be
R_0 = \beta_P / \kappa_P +\phi_M \beta_M / \kappa_M  + (1-\phi_M)[\beta_C/\kappa_C+\phi_{C} \beta_C/\kappa_{H}] \ .
\label{R0eff}
\te
$R_0$ is then a weighted average over the three possible paths connecting $S$ and $R$.

The values of the parameters $\kappa$ are taken from~\cite{Mahajan2020.06.21.20136580}, in units of inverse days:
\begin{equation}
	\begin{aligned}
&\kappa_E=1/(3.1 d)\ , \qquad \kappa_P=1/(2 d)\ , \qquad \kappa_M=1/(13 d)\ ,\\
&\kappa_C=1/(2 d)\ , \qquad \kappa_{C'}=1/(11 d)\ , \qquad \kappa_{H}=1/(11 d)\ ,
	\end{aligned}
	\label{kappa}
\end{equation}
we elaborate more on these times in Appendix~\ref{appendix-fluctuations}.

We select two benchmark points which exemplify admissible combinations of parameters:
\begin{equation}
\begin{aligned}
	& {\rm BP1}:\ \phi_M=0.5,\qquad \phi_C=0.3,\qquad  \beta_P=\beta_M=2\beta_C \ \\
	& {\rm BP2}:\ \phi_M=0.5,\qquad \phi_C=0.3,\qquad  \beta_P=4\beta_M=2\beta_C \ .
\end{aligned}
	\label{BP}
\end{equation}
Refs.~\cite{Mahajan2020.06.21.20136580,oxfordModel} have shown that these are plausible values that can describe the epidemics in different countries.  When considering these benchmark points, the values of $\beta_{P,M,C}$  vary to describe different levels of lock-down, but the ratios are kept fixed.   In addition to the above, as a simplified pedagogical example we also include a benchmark point with $\phi_M=0$ and $\phi_C=0$
\begin{equation}
	\begin{aligned}
&{\rm BP0}:\ \phi_M=0,\qquad \phi_C=0,\qquad  \beta_P=\beta_M=2\beta_C 
\end{aligned}
\label{BPp}
\end{equation}

\subsection{Coupled model in a many blocks system}
We implement the model in a system of $\mathcal{N}$ blocks that are connected by fluxes. The details can be found in Appendix \ref{section:appendix_interconnected}. The population flux can accommodate relevant features of the COVID-19 epidemic, such as for instance a second wave, which is not present in a single isolated block. A similar treatment has been implemented in~\cite{CONTRERAS2020109925}.

We consider that a fraction of the population of each block travels during the day to other blocks and returns. We make the approximation that the traveling population only interact in the visiting block before returning to their original block. By interaction we mean that each person has the health conditions of their original block, and interacts with persons in the visiting block according to the health policies in it.  This approximation of interacting only in the visiting block is made in order to simplify the flux matrix. Relaxing this approximation is no more complicated than defining an smaller time step and a flux matrix for each time of the day considered. As we treat the population on an statistical level, this approximation is not too important to the qualitative behavior of the epidemic.  An important implementation we make when coupling the blocks together is that of an epidemiological barrier, where only non symptomatic people can travel between blocks. Taking this into account, we implement a flux between the $\mathcal{N}$ blocks as a matrix of dimension $\mathcal{N}$x$\mathcal{N}$ where the matrix element $F_{ij}$ is the amount of people from the $S,E,P,M,R$ populations of block $j$ spending their day in block $i$.

When considering $\mathcal{N}$ blocks, each block has its own set of parameters $\beta,\kappa,\phi$, however, the purely biological parameters $\kappa$ do not change between blocks. The main difference between blocks will be on the parameters which depend on societal factors. In particular, we encode all these societal differences in the infection parameters $\beta_{P,M,C}$ and set $\phi_{M}$ and $\phi_{C}$ to be the same over all $\mathcal{N}$ blocks. This is equivalent to saying that we expect the testing capabilities and the individual responsibility to be mostly the same in all the blocks. 
When an individual from block $j$ travels to block $i$, it adopts the relevant local $\beta_{i}$ to interact with the current population of block $i$.

With this approximation, we have a $\mathcal{N}$ block system with $\mathcal{N}(\mathcal{N}-1)$ parameters for the population flux and $3\mathcal{N}+8$ parameters for the disease modeling. These are implemented in a generalized set of equations which take into account the movement of interacting subjects. The algorithm is the following:
\begin{itemize}
\item At the beginning of the day, a given fraction of the $S,E,P,M, R$ compartment populations move from their own block to others.
\item The new populations in each block evolve according to Eqs.~(\ref{slaq_2}) and the stochastic method described in Appendix~\ref{appendix-fluctuations}.
\item The changed populations travel back to their own block.
\end{itemize} 

Technical details of the description of $\mathcal{N}$ blocks individually obeying Eqs.~(\ref{slaq_2}) and coupled through a flux matrix can be found in Appendix~\ref{section:appendix_interconnected}.
\section{The firewall block}
\label{section:two}

The goal of this work is to show that surrounding a COVID-19 outbreak in a city with a firewall block, yields a strong suppression in the probability of spreading the epidemic to the rest of the city.    For the sake of posing the problem, we first consider the configuration where there are only two blocks without a firewall: one for the outbreak and another for the rest of the city.  We then study the configuration with and without firewall and compare the probabilities of having a widespread phase in the bulk of the city as a function of the variables of the problem.  

The setup for a city divided in two blocks consists in having an outbreak in a small block (S) and none infected nor exposed people in the large block (L). For the sake of concreteness we consider a 1M people city divided as follows:
\begin{equation}
        \begin{array}{ccc}
        \mbox{small block (S)} & \longleftrightarrow  & \mbox{large block (L)} \\
        \mbox{50k} && \mbox{950k}
        \end{array}
        \nonumber
\end{equation}
We assume that people crossing the border between these two blocks is controlled in number and that only asymptomatic persons are allowed to cross in any direction.   The model that we have defined has many parameters, however we find that two of them have the largest impact on the evolution of the system: $F_{ij}$ and $R_0^{(i)}$, where $i,j=S,L$. For the sake of simplicity we take along this section symmetric fluxes $F_{ij} = F_{ji}$.  Since we want to study the impact of an outbreak on the rest of the population, from now on we fix $R_0^{(S)}=0.9$, assuming that a lock-down is active on block S, and we take $E_S=500$ and $E_L=0$ at $t=0$. We simulate the evolution of the system while varying $F_{SL}$ and $\beta_P^{(L)}$, by considering the benchmark points of Eq.~(\ref{BP}).

\begin{figure}[h!]
	\begin{center}
	\subfloat[]{\includegraphics[width=0.7\textwidth]{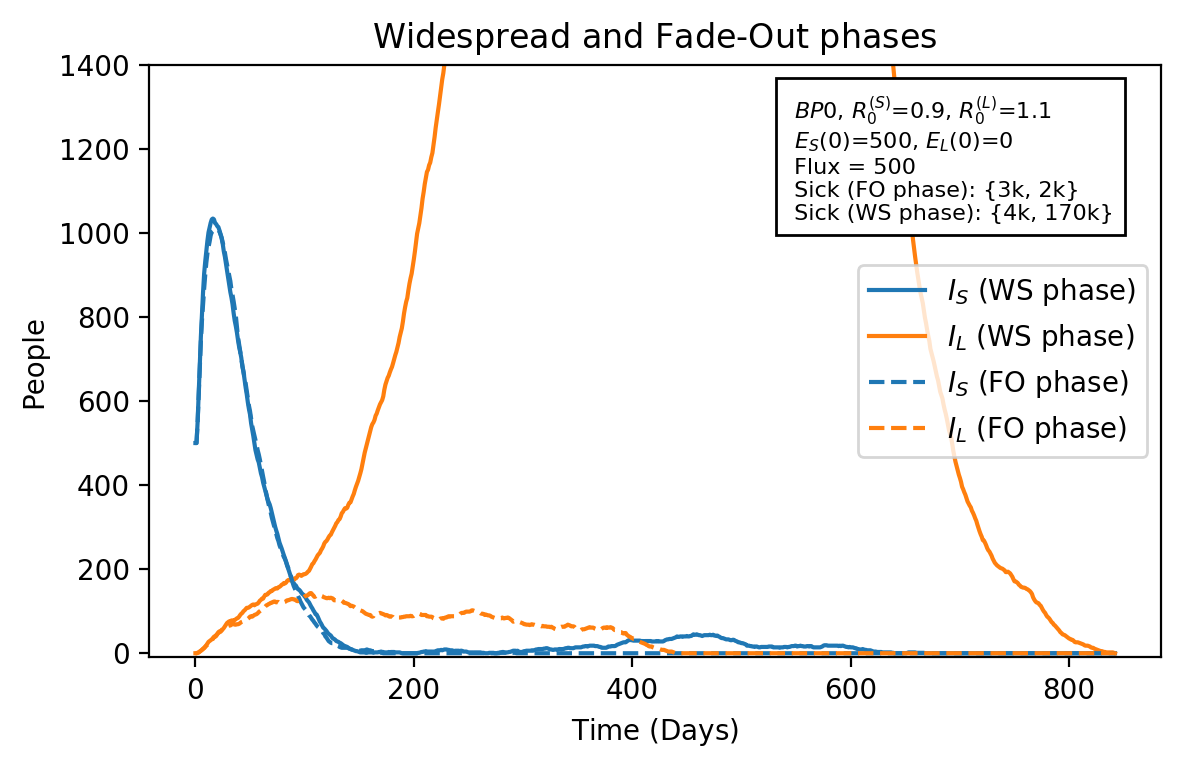}}
	\end{center}
	\caption{
		Time evolution of infected populations for a configuration of two blocks, one small (S) with 50k people and an outbreak of 500, the other large (L) with 950k people and neither infected at $t=0$, and a flux between them of $F_{SL}=F_{LS}=500$. The plot shows $I_S$ and $I_L$ as function of time for two simulations with the same parameters and initial conditions, with $R_0^{(L)}=1.1$, slightly larger than the critical value of 1. In the first simulation (dashed) the epidemic is fade-out (FO), whereas in the second one (continuous) the epidemic is widespread (WS) in block L.}
\label{2-blocks-fluctuations}
\end{figure}

As stated in Section \ref{section:model}, the evolution is not deterministic due to the introduction of noise, thus the same initial conditions and parameters may yield different outcomes of the system.  In particular, considering the above setup there are two well distinguishable phases for the outcome of block L: widespread (WS) and fade out (FO). We can visualize this by analyzing the time evolution of the system in Fig.~\ref{2-blocks-fluctuations}, where we plot the total number of infected people on each block as function of time: $I_i=E_i+P_i+C_i+C'_i+M_i+H_i$, for two simulations with the same parameters and initial conditions, taking BP0. In the first simulation, plotted with dashed lines, only a small number of people is infected on block L. In that case we obtain $P_L(t)+C_L(t)\lesssim {\cal O}(10)$ at all times, 
with the city evolving to the FO phase. On the other hand, in the second simulation plotted with continuous lines, a large number of people is infected on block L, leading to the WS phase. Both simulations have been run with the same parameters and the different results are only due to the implemented stochastic noise and to the fact that $R_0^{(L)} \gtrsim 1$. If the flux is large, such that several people are infected on block L, the epidemic is triggered, if the flux is very small, such that in average no infected people enters on block L, the epidemic is not transmitted. In the middle there is a region of transition, where there is a given probability that the epidemic is triggered on block L. 

\begin{figure}[h!]
\centering
\hskip 2.4cm \includegraphics[width=0.7\textwidth]{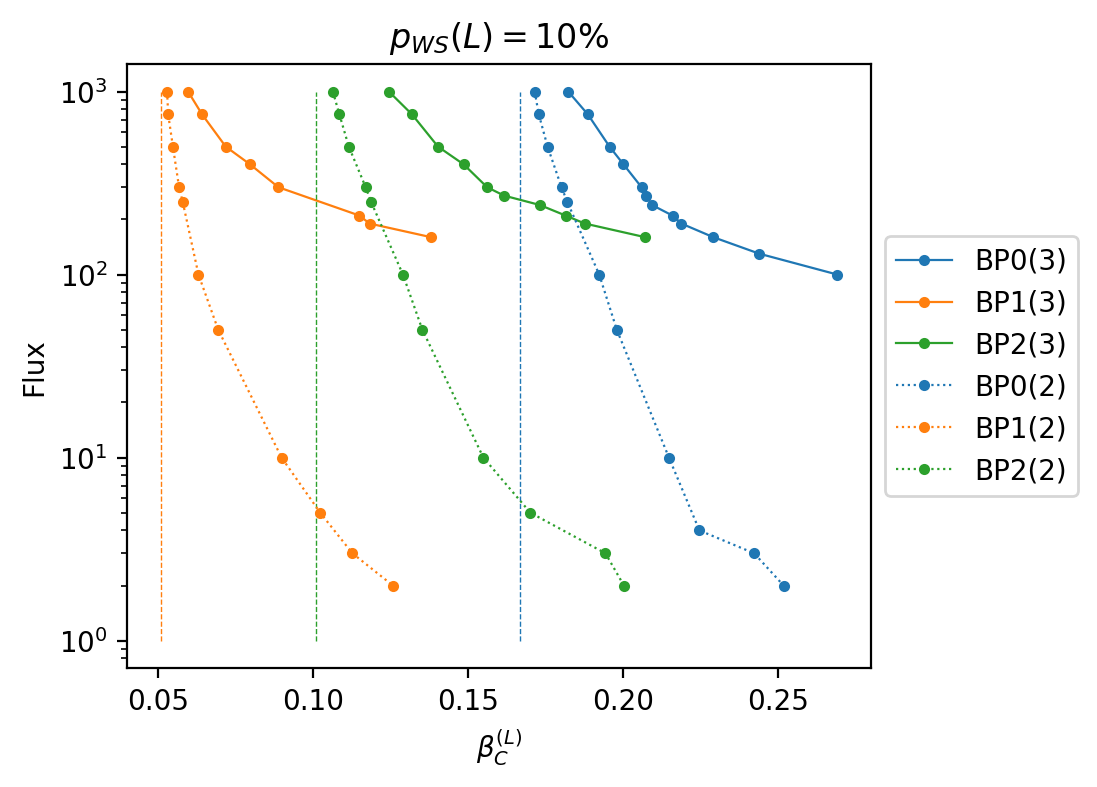}
	\caption{
		Contour curves for 10\% probability of widespread phase on the large block (L) for two blocks setup (dashed lines) and three blocks setup (solid lines).  In both cases there is a small block (S) with an outbreak of 500 infected people, and a large block (L) with no infected, but in the three blocks setup there is in addition a firewall block (F) in the middle that acts as a buffer.  Vertical lines correspond to $R_0=1$ for each benchmark point and the $\beta^{(L)}_C$ in the horizontal axis.  The comparison of both curves shows the convenience of introducing the firewall block F.  More details in the text.
		}
\label{buffer}
\end{figure}

By running a large number of simulations, $N_{SIM}$, with the same parameters and initial conditions, one can approximate the probability for the $WS$ phase as: $p_{\rm WS}=N_{WS}/N_{SIM}$, with $N_{WS}$ the number of simulations leading to phase $WS$.
We have determined the $10\%$ probability contours of triggering the epidemic on block L: $p_{\rm WS}=0.1$. We show this probability with dotted lines in Fig.~\ref{buffer} as function of the flux on the vertical axis and $\beta_C^{(L)}$ on the horizontal axis. 
The different dotted curves correspond to different benchmark points and the vertical dotted lines show $R_0^{(L)}=1$. 
The large dots are the calculated points, the oscillations and discontinuities in the derivative of the curves are likely due to limited CPU computation time. On the upper right region of the curves the epidemic is triggered and widely spread on block L, whereas on the lower left region it is not. As expected, for fixed $\beta_C^{(L)}$, the effect of $\phi_M$ and $\phi_C$ is to lower the critical flux, with a stronger dependence on $\phi_M$. 

As it can be seen from these results, unless the  value of $R_0^{(L)}$ is small, the flux has to be extremely reduced to avoid the transmission of the disease.  This might be one of the reasons why there is practically no success in controlling the epidemic from being transmitted from an outbreak with permeable barriers.

Once the epidemic is triggered on block L by transmission from block S, since $N_L\gg N_S$ and $R_0^{(S)}<1$, block L evolves driven by its own dynamics, that is mostly captured by $R_0^{(L)}$. In particular, if all the parameters remain constant in time, the duration and total number of infected people can be understood by studying one block with these parameters.

Given the above results for a configuration of two blocks, we study in the following a setup consisting in three blocks connected in series as a way of suppressing the transmission of the epidemic:
\begin{equation}
	\begin{array}{ccccc}
		\mbox{small block (S)} & \longleftrightarrow & \mbox{firewall block (F)} & \longleftrightarrow & \mbox{large block (L)}  \\
		\mbox{50k} && \mbox{50k} && \mbox{900k}
	\end{array}
	\nonumber
\end{equation}
Analogously to the previous configuration, we set the initial condition of an outbreak going on in block S with $E_S=500$ and no exposed in blocks F and L.  The rules of the system are such that people can only cross one border and return to their original block, people can not cross two borders connecting blocks S and L. In comparison to the previous configuration, we can visualize that we have added block F surrounding block S to control the infection spread from S to L.  

The main idea in this setup is to explore qualitatively whether adding a firewall block around an outbreak can help to control the propagation to the rest of the city.  The objective is to study if this could happen while the rest of the city is having a more relaxed behavior (larger $\beta^{(L)}_C$) than the small (S) and firewall (F) blocks.  We first study how block F with a fixed setup affects the dynamics of the system by just acting as a buffer.  We then explore different scenarios in which the evolution in the number of infected could trigger modifications in the system setup to control the propagation in the large block.

In the first scenario, we implement the firewall block F and study upon which conditions there would be a probability larger than 10\% of having a widespread phase in block L.  This is similar to the previous calculation, but with the new firewall block in between the outbreak in block S and block L.  We perform this study as a function of $\beta^{(L)}_C$ in block L and the symmetric flux allowed between blocks $S\leftrightarrow F$ and $F\leftrightarrow L$ which, for simplicity, we consider all equal.  Results for the three Benchmark Points BP0, BP1 and BP2, and reproduction numbers $R_0^{(S)}=0.9$ and $R_0^{(F)}=1.05$, are shown in Fig.~\ref{buffer} as continuous lines.   As it can be seen, for the same $\beta^{(L)}_C$ conditions in the large block, the firewall acting as a buffer has a threshold flux which is one to two orders of magnitude larger than the case with no firewall.  This is the main result in this work.  One could also read Fig.~\ref{buffer} as follows: for the same flux and $\beta^{(L)}_C$ conditions, the firewall reduces the probability of a widespread phase in the large block.   

The reason for this behavior is that since only asymptomatic persons are allowed to cross the borders, and people can only spend the day in their contiguous block, then it is less likely that exposed people reach block L.  The implementation of this firewall block in a real world scenario would require, for instance, that supplies carried in a transport from block L to block S switch driver in block F after correctly sanitizing the transport.  The reason behind this is that objects can be instantaneously sanitized, whereas there is no way of instantaneously sanitizing people even if they do not show symptoms.

\begin{figure}[h!]
\centering
	\subfloat[]{\includegraphics[width=0.5\textwidth]{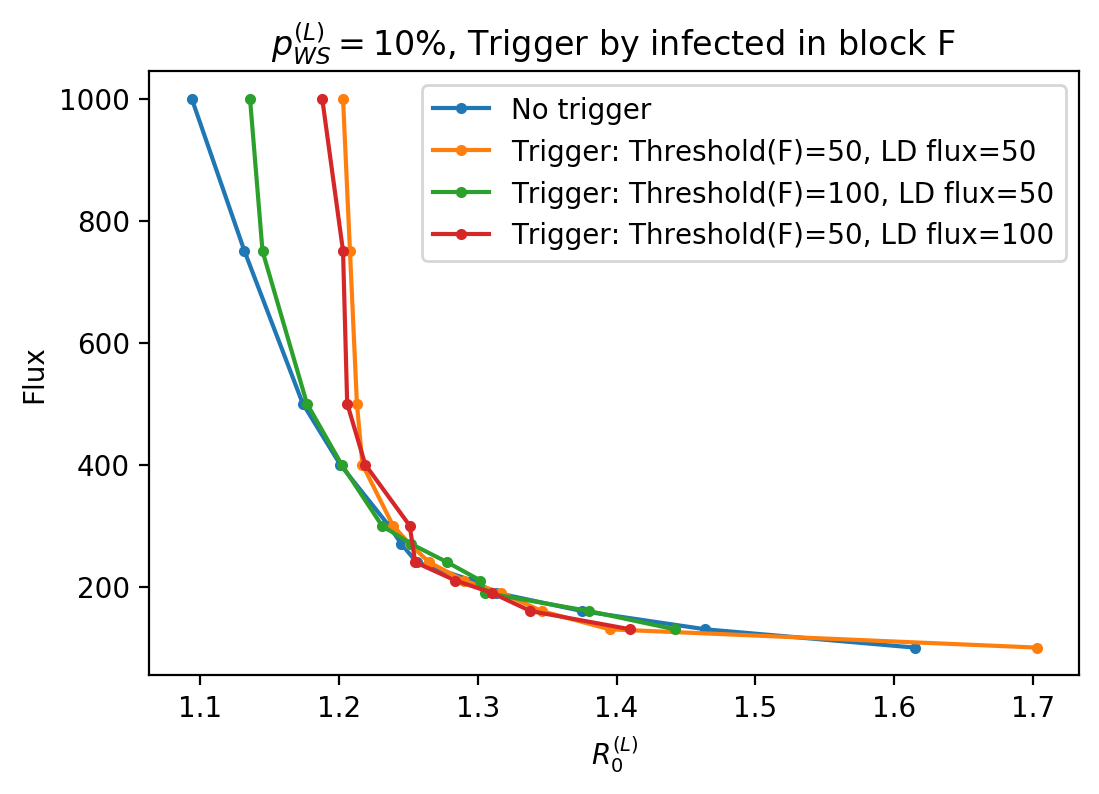}}
	\subfloat[]{\includegraphics[width=0.5\textwidth]{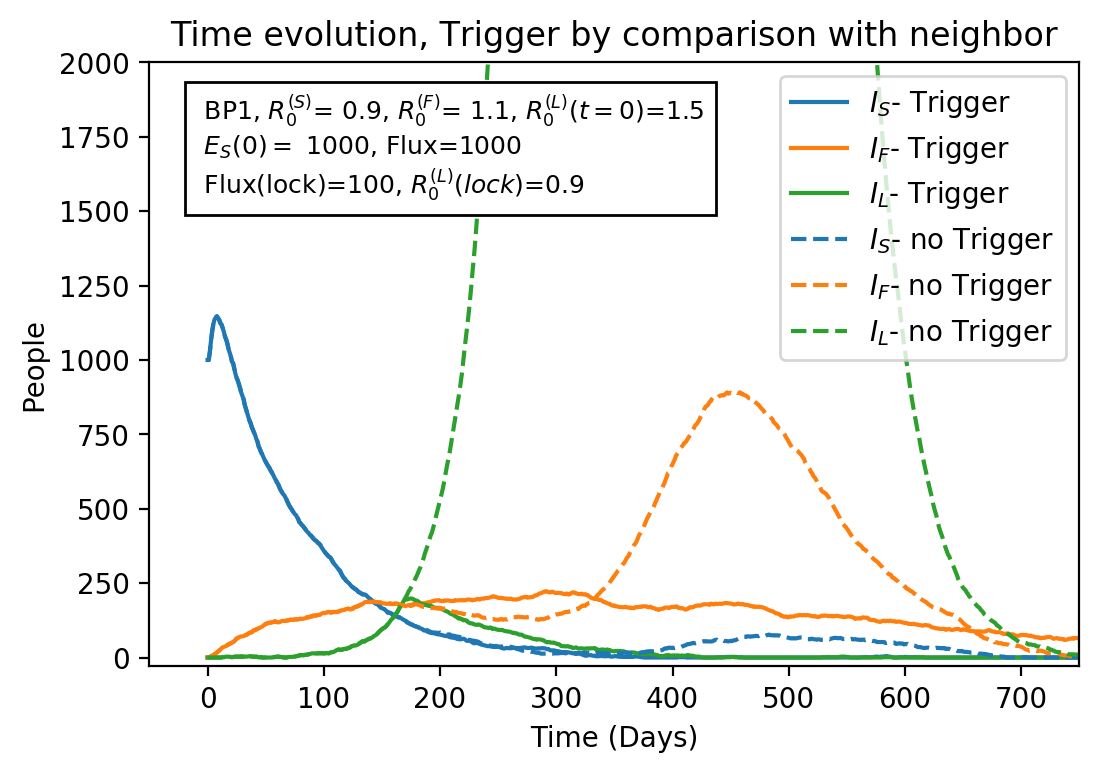}}
	\caption{Two particular cases of a configuration with three blocks and firewall setup in which a given condition triggers a change in the rules.  In the left plot, when the number of infected people in the firewall block (F) surpasses a given threshold, then all fluxes are reduced as stated in legend box, in this case we show only BP0.   In the right plot, using BP1, we explore a scenario in which the large block (L) has a relaxed situation ($R_0^{(L)}=1.5$) until the number of infected in block L surpasses those in block F.  This condition triggers a reduction in the reproduction number $R_0^{(L)}\to 0.9$ and all fluxes to 100 people/day.   Dashed lines show the evolution if the change of rules would have not been triggered.  As expected, a drastic reduction in the reproduction number in block L below 1 controls the widespread of the epidemic.}
\label{firewall}
\end{figure}

In a second stage, we study how this scenario could be improved if, by taking advantage of the firewall block, the city can trigger special policies upon the evolution of the system.  There are many possible triggers to study, for the sake of concreteness we focus on two of them.  In a first case, we consider that if the number of infected people in the firewall block exceeds a given threshold, then all fluxes are reduced to a given minimum.   In Fig.~\ref{firewall}a we show for this setup the same 10\% probability of widespread phase contour curves as in Fig.~\ref{buffer}, while varying the triggering conditions for BP0 and keeping the previous reproduction numbers.  This case, shows the clear convenience of having a triggering firewall for this benchmark point.  However, in other benchmark points with $\phi_M=0.5$ the improvement is negligible.  The reason is that even if the flux is restricted, the people with mild symptoms would be still crossing and then would be equally likely to have the widespread phase in block L.  In a second case, we explore different initial conditions, trigger and change of rules.  We analyze a case in which reproduction numbers before triggering are $R_0^{(S)}=0.9$, $R_0^{(F)}=1.1$ and $R_0^{(L)}=1.5$.   We define to trigger the change in rules when the number of infected in block L surpasses the number of infected in the firewall block F.  This is an important trigger because in this setup block L has a considerable larger reproduction number than the firewall block F.  The change of rules upon triggering conditions is to reduce all fluxes to 100 people/day and $R_0^{(L)} \to 0.9$, which is a drastic change in behavior for the large block.  In Fig.~\ref{firewall}b we show the time evolution for this setup with BP1.

\section{Conclusion}
\label{section:conclusion}
We have studied the spread of an epidemic as COVID-19 for a city divided in blocks, between which only a controlled flux of non-symptomatic people is allowed. In this proposed scenario blocks are artificial divisions which may be based on existing natural or artificial separations, or created ad-hoc. 

We have considered a compartment model to describe the evolution of the epidemic and we have added noise to the system to include stochastic processes. As a result, for $R_0 \gtrsim 1$ the outcome of the differential equations is well divided into two phases, namely fade-out and widespread phases. Given the difficulty to make quantitative predictions, we have studied some generic properties at a qualitative level.

In particular we have investigated the case of an outbreak in a specific neighborhood, which is then separated from the rest of the city by a epidemiological barrier. We have estimated the probability of having a widespread phase in the bulk of the city as function of the flux between the blocks, the compartment contagion factors and the reproduction number. We obtained that, above the critical value $R_0=1$, the $10\%$ probability of a widespread reaches very small fluxes with mild changes of $R_0$.

We have also considered the addition of an intermediate firewall block between the other two, forbidding individuals to pass from the outbreak to the bulk. We found a considerable improvement in controlling the epidemic by just using it as a buffer, allowing fluxes of one to two orders of magnitude larger than in the case of no firewall. We have also studied options of looking at specific variables which trigger flux and lock-down modifications in the system, but we did not found relevant improvements beyond the firewall as a buffer. Further studies in this direction may be fruitful.

The implementation of barriers has many complex social and logistic counterparts, which are very important, that we have not considered.  The aim of this work was to provide the mathematical side of the modeling, as well as qualitative results of the proposed scenario, as one of the necessary tools for an eventual political decision.

At last, it is worth observing that there are equivalent situations to which the results obtained in this work could be translated.  One of them is that, instead of considering blocks within a city, one could apply these results to a set of regions, countries, or towns, as long as similar conditions hold.   In a similar manner it could happen that, instead of having an outbreak, one has a specific region with no infected people in it.  In such a case, implementing a firewall around it would also result in an enhanced protection from the surrounding regions.

\section*{Acknowledgments}
We thank Marcelo Kuperman for very fruitful discussions and for reading an early version of this article, L.D. also thanks Fabiana Laguna and Dami\'an Zanette for discussions and Centro de Estudios Patag\'onicos for support.
\appendix
\section*{Appendices}

\section{Stochastic implementation of the model}
\label{appendix-fluctuations}

In this appendix we detail how Eqs.~(\ref{slaq_2}) are solved in a stochastic manner by introducing several sources of fluctuations.

The system of Eqs.~(\ref{slaq_2}) is deterministic when the populations are continuous functions of time and when the values of the parameters $\beta, \kappa, \phi$ have determined values. This is valid when treating the disease as a population-level phenomenon, where individual cases do not matter. However, in some cases the individual behavior can be important, particularly at the onset of the disease, where most of the compartments are scarcely inhabited. This indicates that, to reflect more accurately the evolution of the disease, some degree of non-deterministic behavior must be incorporated into the model. In particular, given the same initial conditions, it should be possible to reach different final states, especially in cases where the disease will most likely stop propagating before infecting a large fraction of the susceptible population. 

We incorporate this probabilistic aspect in a two-fold way. The first step is incorporated naturally when solving the differential equations numerically. There, as detailed in sec.~\ref{section:model}, we set a time-step of one day and compute the change of the different populations according to a discretized version of Eqs.~(\ref{slaq_2}). When discretizing the set of equations, besides discretizing the time step, we take into account that the populations must be integer numbers and so the changes, often non-integer real numbers, must be transformed into integers. This can be done in a deterministic way by rounding or flooring. This transformation, although deterministic, can prevent a change in a given population which would happen otherwise, as we need the population change to be at least 0.5 (1.0) in magnitude when rounding (flooring) for it to survive. To avoid this, we introduce here a stochastic rounding, which acts as a source of fluctuations to the population change. This stochastic rounding acts on a quantity $n+x$, with $x$ its decimal part and $n$ its integer part, as follows. By drawing a real number $y$ distributed uniformly between 0 and 1, if $y\leq x$, the result of rounding $n+x$ is $n+1$, else the result is $n$. This essentially means that the decimal part of the number acts as the probability for the result of the rounding to be one plus the integer part of the number. 

The second source of stochastic behavior appears when calculating the population changes for each compartment. For a well defined set of parameters $\beta$ and $\kappa$, the change is deterministic, however these parameters are statistical in nature and are misleading when applied to individuals. Instead we use these statistical parameters to sample from probability distributions, where the number of samples is equal to the size of each compartment, and we consider the average of these samples. This procedure makes small samples more prone to fluctuations and large samples more deterministic, as we intend to. For the $\beta$ parameters, instead of calculating $\beta_X X$ with a fixed value for $\beta_X$, we sample $\beta_X \sim \text{Norm}(\beta_X^0,\sigma_X)$, $X$ times. The sum of all these numbers is distributed around $\beta_X^0 X$, with a dispersion of $\sqrt{X}\sigma_X$. By doing this we allow for increments to have different values, especially in the limit of small numbers. The other set of parameters that appear in the equations are the decay rates $\kappa_Y$ for each species. For each term of the form $\kappa_Y Y$, we sample $Y$ values from an exponential distribution with scale parameter $\tau_Y$:
\be
t_Y \sim \frac{e^{-\frac{t}{\tau_Y}}}{\tau_Y}
\te
And take the mean of these values as the inverse of the decay rate, that is $\kappa_Y = \frac{1}{<t_Y>}$.  The scale of these fluctuations is constrained by the exponential distribution having the same deviation.

Doing the above amounts to having noise added to central values of the parameters, thus having a different set of values at each step of the time evolution. Along this work we take $\sigma_X$ to be sufficiently small for the $\beta$ fluctuations not to play any significant role.

The addition of stochastic fluctuations in the model allows for extinctions of the disease to occur, in cases where in a deterministic model they do not. For instance, for a basic reproduction number larger than 1, in a deterministic model there is always an exponential growth of the epidemic. However, for a stochastic model, there can be realizations in which the disease fades out before being able to infect a sizeable part of the population. This is usually referred to as noise-induced phase transitions.

The code is implemented in {\tt Python 3} and can be made available upon request. 

\section{Interconnected blocks}
\label{section:appendix_interconnected}
In this appendix we describe how the extension of the model to include movement between different blocks is done. We consider that a population is divided into several locations, each one being described by the quantities of Eqs.~(\ref{slaq_2}), that we refer to with a block index, $S_i,P_i,{\rm etc}$. Now we want to modify this set of equations to take into account how individuals can move from one block to another during the day and in this way infect people from places different than those they reside in. This represents people commuting to work in a part of the city that is different from the one they live in. Doing this would amount to allowing the $\beta$ coefficients in the above model to have two sets of indices, representing both initial and final blocks, i.e. $\beta^{ij}$. However, we wish to also account for healthy individuals, traveling away from their own block, that may get infected at a different block. It is for this reason that we find better to parameterize new contagions in the following way. We consider having infection rates at each block, of each kind $\beta_P^i,\beta_M^i,\beta_C^i$. In addition, we parameterize the movement of people between different blocks by using a Flux matrix $F_{ij}$. This matrix has certain properties and constraints to obey. $F_{ij}$ represents people from block $j$ that are currently at block $i$. For this reason, for a specific moment, the sum over a certain column of this matrix has to be equal to the total amount of ``available people`` originating from that block. This means
\be
\sum_{j} F_{ji} = N_i - H_i  
\te
Here we discretize the time evolution in the equations, into day-long steps. This means that rates are measured in units of ${\rm day}^{-1}$. This is done in order to have a well defined meaning for the flux matrix, and to calculate the new infected per day as a function of the flux matrix, and the rates of infection per block. We consider then, that the flux matrix represents on a single day the amount of people from each of the blocks that spend the day in another block. We can thus calculate, for a specific target block $i$, how many people are present on a single day, as the sum over a row of the flux matrix,
\be
\hat{N}_i = \sum_{l} F_{il}
\te 
In the same way, we can use this matrix to calculate how many contagious and susceptible people from every block can be present at target block $i$ on a given day. As we are assuming that there are epidemiological barriers between blocks, these keep contagious people from traveling between blocks. Basically, we do not allow for  people from compartments $C$ and $C'$ to leave their own blocks. With these restrictions, the way the disease can go from one block with the disease to another block is either with the travel of pre-symptomatic individuals $P$, people with a mild or asymptomatic version of the disease $M$, or with susceptible people acquiring the disease and bringing it back to their original block. These restrictions on travel translate into constraints on the allowed values in the flux matrix. We can see this acts as a bound on the maximum amount of people leaving a certain block into different ones
\be
\max\left( \sum_{j\neq i} F_{ji} \right) = S_i + E_i + P_i + R_i + M_i = N_i - H_i - C_i -C'_i
\te
or as a bound on the minimum value that the diagonal element can have, that is the amount of people remaining in their block of origin:
\be
\min\left(F_{ii}\right) = C_i + C'_i
\te
Now, with this in mind we can calculate the amount of pre-symptomatic individuals coming into target block $i$ from different blocks as
\be 
\sum_{k\neq i} \frac{F_{ik} P_k}{N_k-H_k-C_k-C'_k}
\te
Whereas the number of pre-symptomatic individuals remaining in their block of origin $i$ has a different expression, that can be found by requiring that the total number of pre-symptomatic leaving from block $i$ add up to $P_i$. That is,
\al
P_i^{\rm remain} &= P_i - P_i^{\rm leave}= P_i - \sum_{j\neq i} \frac{F_{ji} P_i}{N_i-H_i-C_i-C'_i} \nn \\
&= P_i \left( 1 -  \frac{\sum_{j} F_{ji} - F_{ii}}{N_i-H_i-C_i-C'_i} \right) = P_i   \frac{ F_{ii} - \left(C_i +C'_i\right)}{N_i-H_i-C_i-C'_i}
\fal 
Which can be shortened by writing $C_i+C'_i={\rm min}\left(F_{ii}\right)$. Using this, we can express the total amount of pre-symptomatic individuals at a given block as
\be
\hat{N}_P^i = \sum_k \frac{F_{ik} P_k  \left(  1-\delta_{ik}  \frac{{\rm min}(F_{ii}) }{F_{ii}} \right) }{N_k-H_k-C_k-C'_k}
\te
A similar expression gives the amount of susceptible and of mild individuals at target block $i$. The number of individuals from compartments $C$ and $C'$ are simply $C_i$ and $C'_i$, as they cannot travel between blocks. The number of new infections happening at this block is then calculated as the first equation Eqs.~(\ref{slaq_2}).
\be
\Delta \hat{N}_E^i = \frac{\hat{N}_S^i}{\hat{N}^i} \left(\beta_P^i \hat{N}_P^i + \beta_M^i \hat{N}_M^i + \beta_C^i (C_i + C'_i) \right)
\te 
This gives the number of new infections occurring at this block $i$. This, however, does not correspond directly to the number that must be added to $E_i$. We have to distribute these infections over all the blocks that contributed with susceptible individuals to this block, and we do so following the proportions of susceptible from each block over the total susceptible at target block. That is, the contribution of new infected to block $n$ happening at block $i$ is
\be
\Delta E_n^{i}= \frac{F_{in} S_n \left( 1- \delta_{in} \frac{{\rm min}(F_{ii})}{F_{ii}}  \right) }{N_n - H_n-C_n-C'_n} \frac{\Delta \hat{N}_E^i}{\hat{N}_S^i}
\te
If we now add over the block where the infections are happening ($i$), we get the total number of daily new infections for people of block $n$ as
\al
&\Delta E_n = \sum_i \Delta E_n^i = \sum_{ik} \frac{F_{in} F_{ik}}{\sum_{l}F_{il}} \left(\frac{S_n  \left( 1- \delta_{in} \frac{{\rm min}(F_{ii})}{F_{ii}}  \right) }{N_n - H_n-C_n-C'_n}\right) \nn \\
& \times\Bigg\lbrace \frac{(\beta_i^P P_k+\beta_i^M M_k)  \left(  1-\delta_{ik}  \frac{{\rm min}(F_{ii}) }{F_{ii}} \right) }{N_k-H_k-C_k-C'_k} + \frac{\delta_{ik} \beta_i^C \left(  C_i + C'_i\right)}{F_{ii}} \Bigg\rbrace
\label{fini3m}
\fal

\bibliographystyle{JHEP}
\bibliography{biblio}
\end{document}